\pdfoutput=1

\documentclass[11pt,twoside,a4paper,cmspaper,final,collab]{cms-tdr}
\def\svnVersion{253051}\def\cmsCernNo{2013-037}\def\cmsCernDate{\today}\def\cmsMessage{Published in Physics Letters B as \href{http://dx.doi.org/10.1016/j.physletb.2014.06.077}{\doi{10.1016/j.physletb.2014.06.077.}}}

\begin{document}\cmsNoteHeader{HIG-14-002}

\hyphenation{had-ron-i-za-tion}
\hyphenation{cal-or-i-me-ter}
\hyphenation{de-vices}

\RCS$Revision: 246108 $
\RCS$HeadURL: svn+ssh://alverson@svn.cern.ch/reps/tdr2/papers/HIG-14-002/trunk/HIG-14-002.tex $
\RCS$Id: HIG-14-002.tex 246108 2014-06-12 18:42:21Z alverson $
\newlength\cmsFigWidth
\ifthenelse{\boolean{cms@external}}{\setlength\cmsFigWidth{0.95\columnwidth}}{\setlength\cmsFigWidth{0.65\textwidth}}
\ifthenelse{\boolean{cms@external}}{\providecommand{\cmsLeft}{top}}{\providecommand{\cmsLeft}{left}}
\ifthenelse{\boolean{cms@external}}{\providecommand{\cmsRight}{bottom}}{\providecommand{\cmsRight}{right}}
\newcommand{\Zee}{\ensuremath{\cPZ\to\Pep\Pem}\xspace}
\newcommand{\tauh}{\ensuremath{\Pgt_\mathrm{h}}}
\newcommand{\taue}{\ensuremath{\Pgt_\Pe}}
\newcommand{\taul}{\ensuremath{\Pgt_\mathrm{\ell}}}
\newcommand{\Hmu}{\ensuremath{\PH\to\cPZ\cPZ^{(\ast)}\to4\Pgm}}
\newcommand{\mmu}{\ensuremath{m_{4\Pgm}}}
\newcommand{\mell}{\ensuremath{m_{4\ell}}}
\newcommand{\mZZ}{\ensuremath{m_{\cPZ\cPZ}}}
\newcommand{\mH}{\ensuremath{m_{\PH}}}
\newcommand{\GH}{\ensuremath{\Gamma_{\PH}}}
\newcommand{\GHs}{\ensuremath{\Gamma_{\PH}^{\mathrm{SM}}}}
\newcommand{\GZ}{\ensuremath{\Gamma_{0}}}
\newcommand{\GHratio}{\ensuremath{\GH/\GHs}}
\newcommand{\lnQ}{\ensuremath{\ln(Q)}}
\newcommand{\mlnQ}{\ensuremath{-2\,\ln(Q)}}
\newcommand{\LS}{\ensuremath{\mathcal{L}(S)}}
\newcommand{\LB}{\ensuremath{\mathcal{L}(B)}}
\newcommand{\LSB}{\ensuremath{\mathcal{L}(S+B)}}
\newcommand{\mZ}{\ensuremath{m_{\cPZ}}}
\newcommand{\Zbb}{\ensuremath{(\cPZ/\Pgg^*){\cPqb\cPaqb}}}
\newcommand{\ZZ}{\ensuremath{\cPZ/\Pgg^*\cPZ/\Pgg^*}}
\newcommand{\Zgam}{\ensuremath{\cPZ/\Pgg^*}}
\newcommand{\mumu}{\ensuremath{\Pgmp\Pgmm}}
\newcommand{\Wo}{\ensuremath{\PW}}%
\newcommand{\Wp}{\ensuremath{\PWp}}%
\newcommand{\Wm}{\ensuremath{\PWm}}%
\newcommand{\Zo}{\ensuremath{\cPZ}}%
\newcommand{\Ho}{\ensuremath{\PH}}%
\newcommand{\UpsNs}{\ensuremath{\PgU(\mathrm{nS})}}
\newcommand{\KD}{\ensuremath{\mathcal{D}^\text{kin}_\text{bkg}} }
\newcommand{\superKD}{\ensuremath{\mathcal{D}_\text{bkg}} }
\newcommand{\spinKD}{\ensuremath{\mathcal{D}_{J^P}} }
\newcommand{\VDj}{\mathrm{\mathcal{D}_\text{jet}} }
\newcommand{\VDu}{{\PT^{4\ell}} }
\newcommand{\MassD}{\mathcal{D}_\mathrm{m} }
\newcommand{\LikMu}{\mathcal{L}_{3D}^{\mu} }
\newcommand{\LikMuZOj}{\mathcal{L}_{3D}^{\mu,\,\text{0/1-jet}}(m_{4\ell},\KD,\VDu) }
\newcommand{\LikMuDj}{\mathcal{L}_{3D}^{\mu,\,\text{dijet}}(m_{4\ell},\KD,\VDj) }
\newcommand{\LikMuTwoD}{\mathcal{L}_{2D}^{\mu} }
\newcommand{\LikMuOneD}{\mathcal{L}_{1D}^{\mu} }
\newcommand{\LikMass}{\mathcal{L}_{3D}^{m,\Gamma} }
\newcommand{\LikMassTwoD}{\mathcal{L}_{2D}^{m,\Gamma} }
\newcommand{\LikMassOneD}{\mathcal{L}_{1D}^{m,\Gamma} }
\newcommand{\LikSpin}{\mathcal{L}_{2D}^{J^P} }
\newcommand{\X}{\ensuremath{\cmsSymbolFace{X}}\xspace}
\newcommand{\sip}{\ensuremath{\text{SIP}_\text{3D}} }
\newcommand{\fakerate}{\ensuremath{f(\ell,\PT^\ell,\vert\eta^\ell\vert)} }
\newcommand{\strength}{\ensuremath{\mu}}
\newcommand{\muV}{\ensuremath{\mu_{\mathrm{VBF},~\mathrm{V\PH}}} }
\newcommand{\muF}{\ensuremath{\mu_{\Pg\Pg\PH,\,\ttbar\PH}} }
\providecommand{\met}{\ETmiss\xspace}
\providecommand{\mt}{\ensuremath{m_\mathrm{T}}}

\newcommand\T{\rule{0pt}{2.6ex}}       
\newcommand\B{\rule[-1.2ex]{0pt}{0pt}} 
\newcommand{\usedLumiA}{5.1\fbinv}
\newcommand{\usedLumiB}{19.7\fbinv}

\cmsNoteHeader{HIG-14-002}
\title{Constraints on the Higgs boson width from off-shell production and decay to Z-boson pairs}

\date{\today}

\abstract{
Constraints are presented on the total width of the recently discovered Higgs boson,
$\Gamma_{\PH}$, using its relative on-shell and off-shell production and decay rates
to a pair of Z bosons, where one Z boson decays to an electron or muon pair, and the other to
an electron, muon, or neutrino pair. The analysis is based on the data collected by the
CMS experiment at the LHC in 2011 and 2012, corresponding to integrated luminosities of
5.1\fbinv at a centre-of-mass energy $\sqrt{s} = 7\TeV$ and 19.7\fbinv at $\sqrt{s} = 8\TeV$.
A simultaneous maximum likelihood fit to the measured kinematic distributions near the resonance
peak and above the Z-boson pair production threshold leads to an upper limit on the Higgs boson
width of $\Gamma_\PH < 22\MeV$ at a 95\% confidence level, which is 5.4 times the
expected value in the standard model at the measured mass of $\mH = 125.6\GeV$.
}

\hypersetup{%
pdfauthor={CMS Collaboration},%
pdftitle={Constraints on the Higgs boson width from off-shell production and decay to Z-boson pairs},%
pdfsubject={CMS},%
pdfkeywords={CMS, physics, higgs, diboson, properties}}

\maketitle 

The discovery of a new boson consistent with the standard model (SM) Higgs boson by the ATLAS and CMS
Collaborations was recently reported~\cite{Aad:2012tfa, Chatrchyan:2012ufa, Chatrchyan:2013lba}.
The mass of the new boson (\mH) was measured to be near 125\GeV, and the spin-parity properties
were further studied by both experiments, favouring the scalar, $\mathrm{J}^{\mathrm{PC}} = 0^{++}$,
hypothesis~\cite{Chatrchyan:2012br,Aad:2013wqa,Aad:2013xqa,Chatrchyan:2013legacy}. The measurements were
found to be consistent with a single narrow resonance, and an upper limit of 3.4\GeV at a 95\% confidence level
(CL) on its decay width (\GH) was reported by the CMS experiment in the four-lepton
decay channel~\cite{Chatrchyan:2013legacy}. A direct width measurement at the resonance peak is
limited by experimental resolution, and is only sensitive to values far larger than the expected
width of around $4\MeV$ for the SM Higgs boson~\cite{LHCHiggsCrossSectionWorkingGroup:2011ti,Heinemeyer:2013tqa}.

It was recently proposed~\cite{CaolaMelnikov:1307.4935} to constrain the Higgs boson width using its
off-shell production and decay to two Z bosons away from the resonance peak~\cite{Kauer:2012hd}.
In the dominant gluon fusion production mode the off-shell production cross section is known to be
sizable. This arises from an enhancement in the decay amplitude from the vicinity of the Z-boson pair production
threshold. A further enhancement comes, in gluon fusion production, from the top-quark
pair production threshold. The zero-width approximation is inadequate and the ratio of the off-shell
cross section above $2m_{\cPZ}$ to the on-shell signal is of the order of 8\%~\cite{Kauer:2012hd,Kauer:1305.2092}.
Further developments to the measurement of the Higgs boson width were proposed in
Refs.~\cite{CampbellEllisWilliams:1311.3589v1,Passarino:1312.2397v1}.

The gluon fusion production cross section depends on \GH~through the Higgs boson propagator
\begin{equation}
\frac{\rd\sigma_{\cPg\cPg \to \PH \to \cPZ\cPZ}}{\rd\mZZ^2}
\sim
\frac{g_{\Pg\Pg\PH}^2g_{\PH\cPZ\cPZ}^2}{(\mZZ^2 - \mH^2)^2 + \mH^2\GH^2},
\end{equation}
where $g_{\Pg\Pg\PH}$ and $g_{\PH\cPZ\cPZ}$ are the couplings of the Higgs boson to gluons and Z bosons,
respectively. Integrating either in a small region around \mH, or above the mass threshold $m_{\cPZ\cPZ} > 2m_\cPZ$,
where $(m_{\cPZ\cPZ} - m_{\PH}) \gg \GH$, the cross sections are, respectively,
\begin{equation}
\label{eq:resonant}
\sigma^\text{on-shell}_{\cPg\cPg \to \PH \to \Zo\Zo^{*}}
\sim
\frac{g_{\Pg\Pg\PH}^2g_{\PH\cPZ\cPZ}^2}{\mH\GH}\;\;\;\text{and}\;\;\;\sigma^\text{off-shell}_{\cPg\cPg
\to \PH^{*} \to \Zo\Zo}
\sim
\frac{g_{\Pg\Pg\PH}^2g_{\PH\cPZ\cPZ}^2}{(2m_\cPZ)^2}.
\end{equation}
From Eq.~(\ref{eq:resonant}), it is clear that a measurement of the relative off-shell and on-shell
production in the $\PH\to\cPZ\cPZ$ channel provides direct information on $\GH$, as long as
the coupling ratios remain unchanged, \ie the gluon fusion production is dominated by the top-quark
loop and there are no new particles contributing. In particular, the on-shell production cross
section is unchanged under a common scaling of the squared product of the couplings and of the total
width $\GH$, while the off-shell production cross section increases linearly
with this scaling factor.

\begin{figure*}[htb]
\centering
\includegraphics[width=0.32\textwidth]{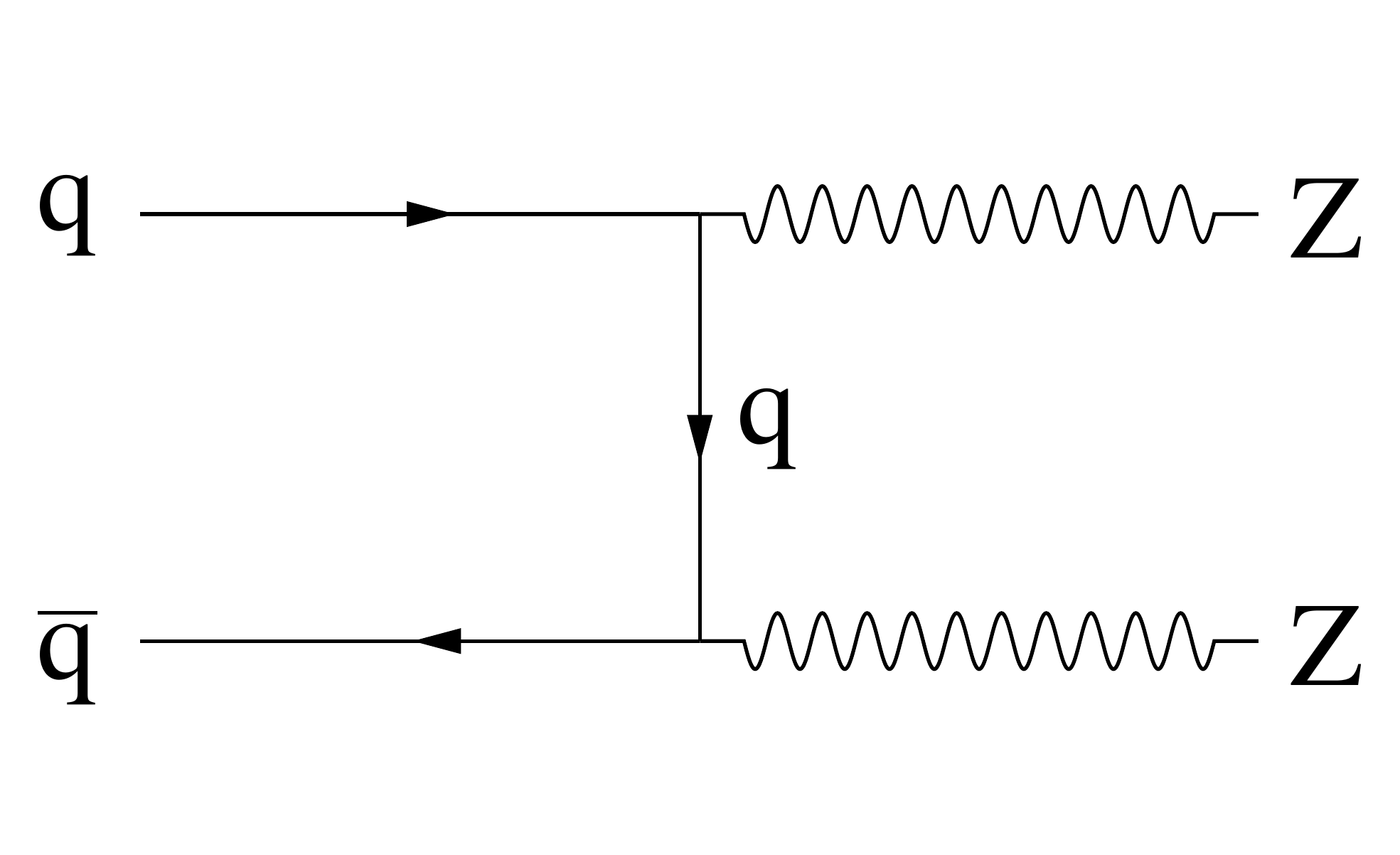}
\includegraphics[width=0.32\textwidth]{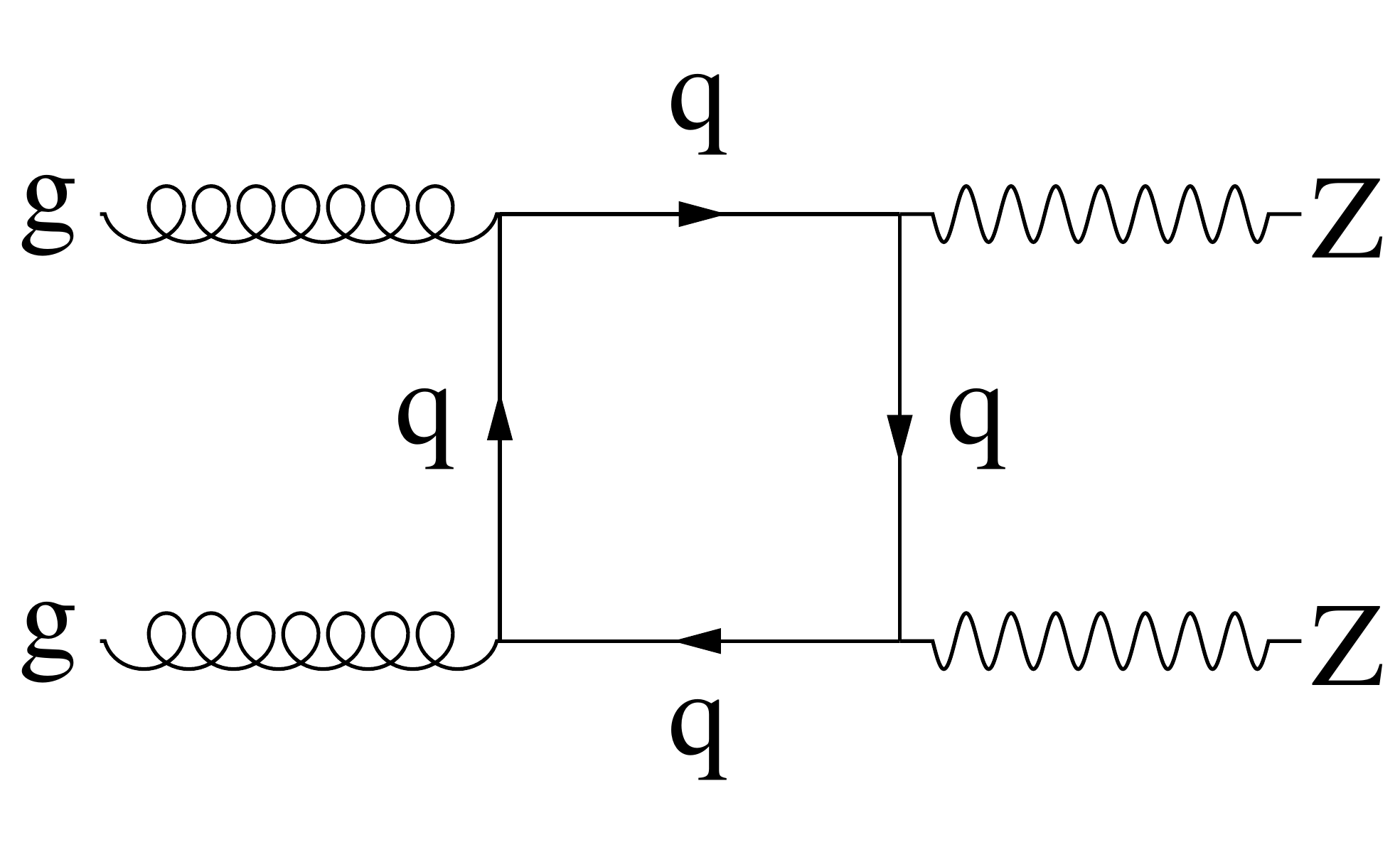}
\includegraphics[width=0.32\textwidth]{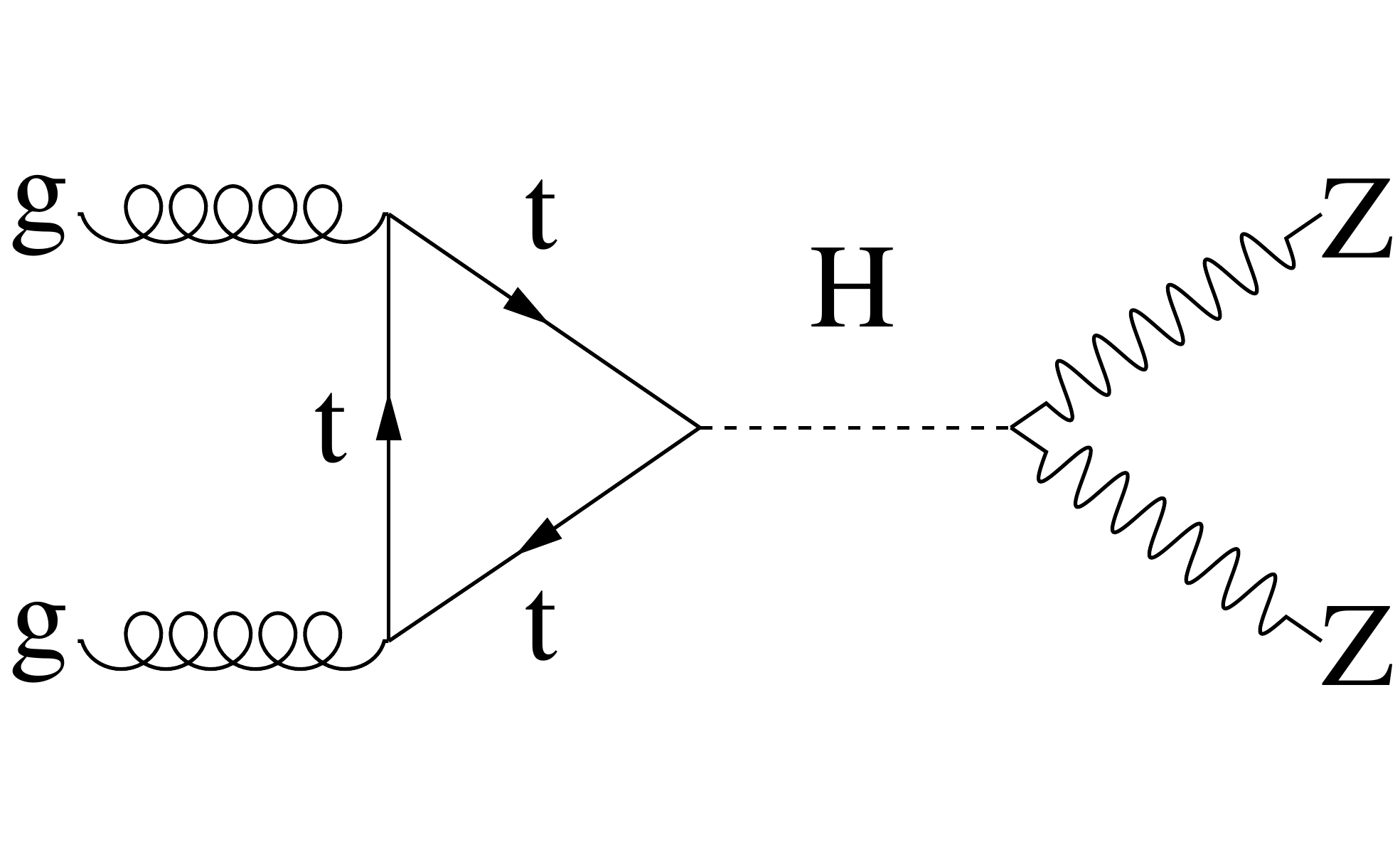}
\caption{
Lowest order contributions to the main $\Zo\Zo$ production processes:
(left) quark-initiated production, $\Pq \Paq \to \Zo\Zo$,
(center) gg continuum background production, $\cPg\cPg \to \Zo\Zo$, and
(right) Higgs-mediated gg production, $\cPg\cPg \to \PH \to \Zo\Zo$, the signal.
}
\label{fig:diagrams}
\end{figure*}

The dominant contribution for the production of a pair of Z bosons comes from the quark-initiated
process, $\Pq \Paq \to \Zo\Zo$, the diagram for which is displayed in Fig.~\ref{fig:diagrams}(left).
The gluon-induced diboson production involves the $\cPg\cPg \to \Zo\Zo$ continuum background
production from the box diagrams, as illustrated in Fig.~\ref{fig:diagrams}(center).
An example of the signal production diagram is shown in Fig.~\ref{fig:diagrams}(right).
The interference between the two gluon-induced contributions is significant at high
$m_{\cPZ\cPZ}$~\cite{Passarino:2012ri}, and is taken into account in the analysis
of the off-shell signal.

Vector boson fusion (VBF) production, which contributes at the level of about 7\% to the on-shell
cross section, is expected to increase above $2m_\cPZ$. The above formalism describing the
ratio of off-shell and on-shell cross sections is applicable to the VBF production mode. In this
analysis we constrain the fraction of VBF production using the properties of the events in the on-shell
region. The other main Higgs boson production mechanisms, $\ttbar\PH$ and VH (V=Z,W), which contribute
at the level of about 5\% to the on-shell signal, are not expected to produce a significant off-shell
contribution as they are suppressed at high mass~\cite{LHCHiggsCrossSectionWorkingGroup:2011ti,Heinemeyer:2013tqa}.
They are therefore neglected in the off-shell analysis.

In this Letter, we present constraints on the Higgs boson width using its off-shell production and decay
to Z-boson pairs, in the final states where one Z boson decays to an electron or a muon pair and the
other to either an electron or a muon pair, $\PH \to \Zo\Zo \to 4\ell$ ($4\ell$ channel), or a pair
of neutrinos, $\PH \to \Zo\Zo \to 2\ell2\nu$ ($2\ell2\nu$ channel). Relying on the observed Higgs boson
signal in the resonance peak region~\cite{Chatrchyan:2013legacy}, the simultaneous measurement of the
signal in the high-mass region leads to constraints on the Higgs boson width \GH~in the $4\ell$ decay
channel. The $2\ell 2\nu$ decay channel, which benefits from a higher branching fraction~\cite{Chatrchyan:2013zz,Chatrchyan:2012llnunu},
is used in the high-mass region to further increase the sensitivity to the Higgs boson width.
The analysis is performed for the tree-level HVV coupling of a scalar Higgs boson, consistent
with our observations~\cite{Chatrchyan:2012br,Chatrchyan:2013legacy},
and implications for the anomalous HVV interactions are discussed. The Higgs boson mass is set to the
measured value in the $4\ell$ decay channel of $\mH = 125.6\GeV$ \cite{Chatrchyan:2013legacy} and the
Higgs boson width is set to the corresponding expected value in the SM of
$\GHs = 4.15\MeV$~\cite{LHCHiggsCrossSectionWorkingGroup:2011ti,Heinemeyer:2013tqa}.

The measurement is based on $\Pp\Pp$ collision data collected with the CMS detector at the LHC in
2011, corresponding to an integrated luminosity of \usedLumiA at the center-of-mass energy of
$\sqrt{s} = 7\TeV$ ($4\ell$ channel), and in 2012, corresponding to an integrated luminosity of
\usedLumiB at $\sqrt{s} = 8\TeV$ ($4\ell$ and $2\ell2\nu$ channels).
The CMS detector, described in detail elsewhere~\cite{CMSDETECTOR}, provides excellent resolution
for the measurement of electron and muon transverse momenta ($\pt$) over a wide range. The signal
candidates are selected using well-identified and isolated prompt leptons. The online selection
and event reconstruction are described elsewhere~\cite{Chatrchyan:2012ufa,Chatrchyan:2013lba,Chatrchyan:2013legacy,Chatrchyan:2013zz}.
The analysis presented here is based on the same event selection as used in Refs.~\cite{Chatrchyan:2013legacy,Chatrchyan:2013zz}.

The analysis in the $4\ell$ channel uses the four-lepton invariant mass distribution as well as a
matrix element likelihood discriminant to separate the ZZ components originating from gluon- and
quark-initiated processes. We define the on-shell signal region as $105.6 < m_{4\ell} < 140.6\GeV$
and the off-shell signal region as  $m_{4\ell} > 220\GeV$. The analysis in the $2\ell 2\nu$ channel
relies on the transverse mass distribution \mt,
\begin{equation}
\mt^{2} = \Bigg[\sqrt{{p_{\mathrm{T},2\ell}}^2 + {m_{2\ell}}^2} + \sqrt{{\MET}^2 +
{m_{2\ell}}^2}\Bigg]^2 - \Bigg[\vec{p}_{T,2\ell} + \vec{E}_\mathrm{T}^\text{miss}\Bigg]^2 ,
\end{equation}
where $p_{\mathrm{T},2\ell}$ and $m_{2\ell}$ are the measured transverse momentum and invariant mass
of the dilepton system, respectively. The missing transverse energy, \MET, is defined as
the magnitude of the transverse momentum imbalance evaluated as the negative of the vectorial sum of
transverse momenta of all the reconstructed particles in the event. In the $2\ell 2\nu$ channel,
the off-shell signal region is defined as $\mt >180\GeV$. The choice of the off-shell regions in
both channels is done prior to looking at the data, based on the expected sensitivity.

Simulated Monte Carlo (MC) samples of $\cPg\cPg \to 4\ell$ and $\cPg\cPg \to 2\ell 2\nu$ events are
generated at leading order (LO) in perturbative quantum chromodynamics (QCD), including the Higgs
boson signal, the continuum background, and the interference contributions using recent versions of
two different MC generators, \textsc{gg2VV} 3.1.5~\cite{Kauer:2012hd,Kauer:1310.7011} and
\MCFM 6.7~\cite{MCFM}, in order to cross-check theoretical inputs. The QCD renormalization
and factorization scales are set to \mZZ/2 (dynamic scales) and MSTW2008 LO parton distribution
functions (PDFs)~\cite{mstw} are used. Higher-order QCD corrections for the gluon fusion signal process
are known to an accuracy of next-to-next-to-leading order (NNLO) and next-to-next-to-leading logarithms
for the total cross section~\cite{LHCHiggsCrossSectionWorkingGroup:2011ti,Heinemeyer:2013tqa} and to
NNLO as a function of $\mZZ$~\cite{Passarino:1312.2397v1}.
These correction factors to the LO cross section (K factors) are typically in the range of 2.0 to 2.5.
After the application of the $\mZZ$-dependent K factors, the event yield is normalized to the cross
section from Refs.~\cite{LHCHiggsCrossSectionWorkingGroup:2011ti, Heinemeyer:2013tqa}.
For the $\cPg\cPg \to \Zo\Zo$ continuum background, although no exact calculation exists beyond LO,
it has been recently shown~\cite{Bonvini:1304.3053} that the soft collinear approximation is able to
describe the background cross section and therefore the interference term at NNLO. Following this
calculation, we assign to the LO background cross section (and, consequently, to the interference
contribution) a K factor equal to that used for the signal~\cite{Passarino:1312.2397v1}. The limited
theoretical knowledge of the background K factor at NNLO is taken into account by including an additional
systematic uncertainty, the impact of which on the measurement is nevertheless small.

Vector boson fusion events are generated with \textsc{phantom}~\cite{Ballestrero:2007}.
Off-shell and interference effects with the nonresonant production are included at LO in these simulations.
The event yield is normalized to the cross section at NNLO QCD and next-to-leading order (NLO)
electroweak (EW)~\cite{LHCHiggsCrossSectionWorkingGroup:2011ti, Heinemeyer:2013tqa} accuracy, with
a normalization factor shown to be independent of $\mZZ$.

In order to parameterize and validate the distributions of all the components for both gluon fusion and
VBF processes, specific simulated samples are also produced that describe only the signal or the continuum background,
as well as several scenarios with scaled couplings and width. For the on-shell analysis, signal events
are generated either with \POWHEG~\cite{powheg,Alioli:2010xd,Bagnaschi:2011tu,Nason:2009ai} production at NLO in QCD and \textsc{JHUGen}~\cite{Gao:2010qx,Bolognesi:2012}
decay (gluon fusion and VBF), or with \PYTHIA 6.4~\cite{Sjostrand:2006za} (VH and
$\ttbar\PH$ production).

In both the $4\ell$ and $2\ell2\nu$ channels the dominant background is $\Pq \Paq \to \Zo\Zo$.
We assume SM production rates for this background, the contribution of which is evaluated by \POWHEG
simulation at NLO in QCD~\cite{powhegZZ}. Next-to-leading order EW calculations~\cite{Bierweiler:1312,Baglio:1307.4331},
which predict negative and $\mZZ$-dependent corrections to the $\Pq \Paq \to \Zo\Zo$
process for on-shell Z-boson pairs, are taken into account.

All simulated events undergo parton showering and hadronization using \PYTHIA.
As is done in Ref.~\cite{Chatrchyan:2013legacy} for LO samples, the parton showering settings
are tuned to approximately reproduce the ZZ $\pt$ spectrum predicted at NNLO for the Higgs boson
production~\cite{hres}. Generated events are then processed with the detailed CMS detector simulation
based on \GEANTfour~\cite{GEANT4-1,GEANT4-2}, and reconstructed using the same
algorithms as used for the observed events.

The final state in the $4\ell$ channel is characterized by four well-identified and isolated leptons
forming two pairs of opposite-sign and same-flavour leptons consistent with two $\cPZ$ bosons.
This channel benefits from a precise reconstruction of all final state leptons and from a very low
instrumental background. The event selection and the reducible background evaluation are performed
following the methods described in Ref.~\cite{Chatrchyan:2013legacy}.
After the selection, the $4\ell$ data sample is dominated by the quark-initiated
$\Pq \Paq \to \cPZ\cPZ \to 4\ell$ ($\Pq \Paq \to 4\ell$)
and $\cPg\cPg \to 4\ell$ productions.

\begin{figure}[tbh]
\centering
\includegraphics[width=\cmsFigWidth]{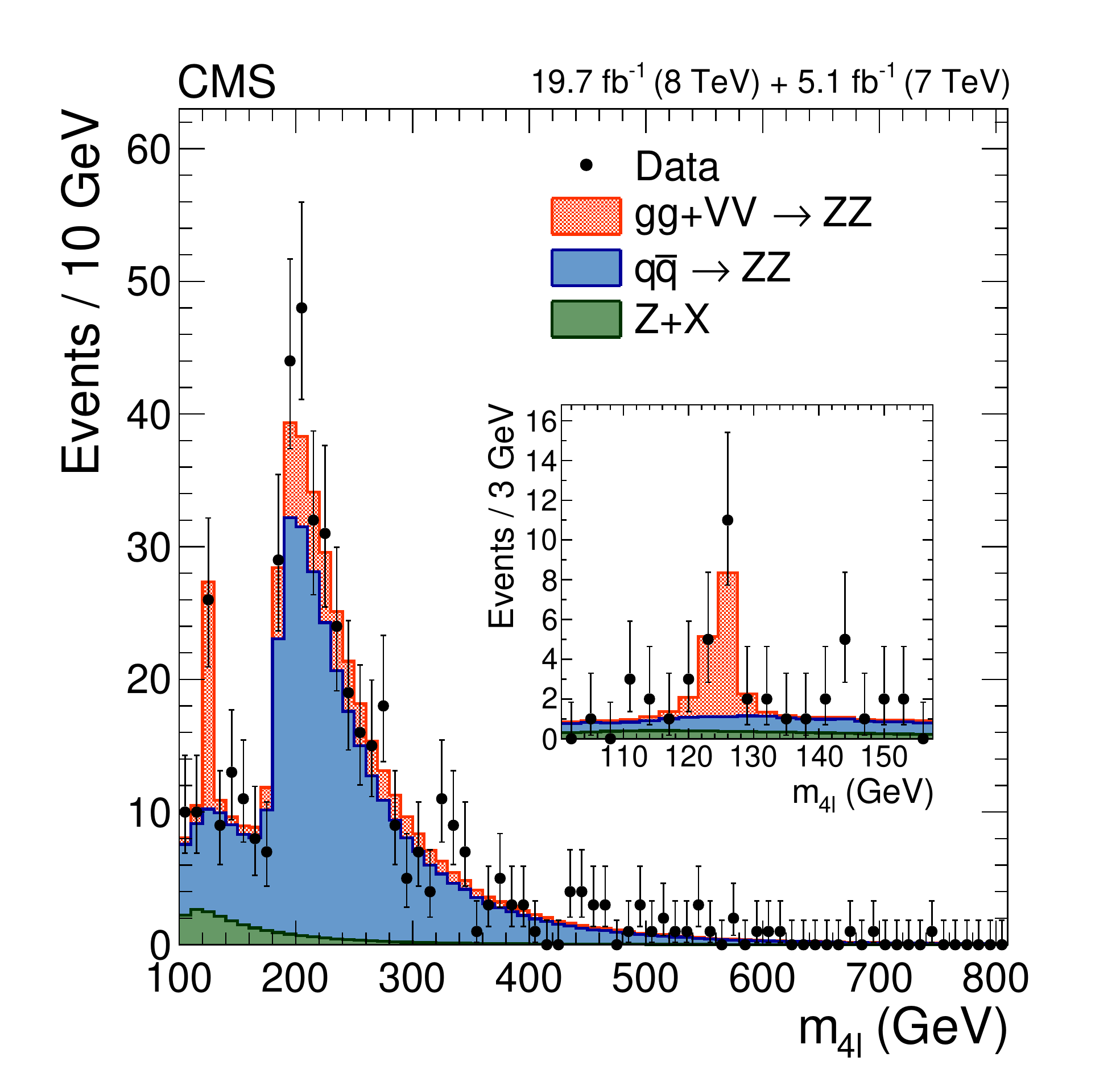}
\caption{
Distribution of the four-lepton invariant mass in the range $100 < m_{4\ell} < 800\GeV$.
Points represent the data, filled histograms the expected contributions from the reducible
(Z+X) and $\Pq\Paq$ backgrounds, and from the sum of the gluon fusion (gg) and
vector boson fusion (VV) processes, including the Higgs boson mediated contributions.
The inset shows the distribution in the low mass region after a selection requirement
on the MELA likelihood discriminant $\mathcal{D}_\text{bkg}^\text{kin} > 0.5$~\cite{Chatrchyan:2013legacy}.
In this region, the contribution of the $\ttbar\PH$ and VH production processes is added to
the dominant gluon fusion and VBF contributions.
}
\label{fig:massfull}
\end{figure}

Figure~\ref{fig:massfull} presents the measured $\mell$ distribution over the full mass range,
$\mell > 100\GeV$, together with the expected SM contributions. The $\cPg\cPg \to 4\ell$ contribution
is clearly visible in the on-shell signal region and at the Z-boson pair production threshold, above
the $\Pq \Paq \to 4\ell$ background. The observed distribution is consistent with
the expectation from SM processes. We observe 223 events in the off-shell signal region, while we
expect $217.6 \pm 9.5$ from SM processes, including the SM Higgs boson signal.

In order to enhance the sensitivity to the gg production in the off-shell region, a likelihood discriminant
$\mathcal{D}_{\cPg\cPg}$ is used, which characterizes the event topology in the $4\ell$ centre-of-mass frame
using the observables $(m_{\cPZ_1}, m_{\cPZ_2}, \vec\Omega)$ for a given value of $\mell$, where
$\vec\Omega$ denotes the five angles defined in Ref.~\cite{Gao:2010qx}. The discriminant is built from the
probabilities $\mathcal{P}^{\cPg\cPg}_\text{tot}$ and $\mathcal{P}^{\Pq \Paq}_\text{bkg}$
for an event to originate from either the $\cPg\cPg \to 4\ell$ or the $\Pq \Paq \to 4\ell$
process. We use the matrix element likelihood approach (MELA)~\cite{Chatrchyan:2012ufa, Bolognesi:2012}
for the probability computation using the \MCFM matrix elements for both $\cPg\cPg \to 4\ell$ and
$\Pq \Paq \to 4\ell$ processes.
The probability $\mathcal{P}^{\cPg\cPg}_\text{tot}$ for the $\cPg\cPg \to 4\ell$ process includes
the signal ($\mathcal{P}^{\cPg\cPg}_\text{sig}$), the background ($\mathcal{P}^{\cPg\cPg}_\text{bkg}$),
and their interference ($\mathcal{P}^{\cPg\cPg}_\text{int}$), as introduced for the discriminant computation in
Ref.~\cite{Anderson:2013afp}. The discriminant is defined as
\begin{equation}
\label{eq:kd-ggmela}
\mathcal{D}_{\cPg\cPg} = \frac{\mathcal{P}^{\cPg\cPg}_\text{tot}  }{\mathcal{P}^{\cPg\cPg}_\text{tot}  + \mathcal{P}^{\Pq \Paq }_\text{bkg} }=
\left[1+\frac{\mathcal{P}^{\Pq \Paq }_\text{bkg}  }
{a \times \mathcal{P}^{\cPg\cPg}_\text{sig} +  \sqrt{a} \times  \mathcal{P}^{\cPg\cPg}_\text{int} + \mathcal{P}^{\cPg\cPg}_\text{bkg}  } \right]^{-1} ,
\end{equation}
where
the parameter $a$ is the strength of the unknown anomalous $\cPg\cPg$ contribution with respect to the
expected SM contribution ($a=1$). We set $a = 10$ in the definition of $\mathcal{D}_{\cPg\cPg}$ according
to the expected sensitivity. Studies show that the expected sensitivity does not change substantially
when $a$ is varied up or down by a factor of 2. It should be stressed that fixing the parameter $a$ to
a given value only affects the sensitivity of the analysis. To suppress the dominant  $\Pq \Paq \to 4\ell$
background in the on-shell region, the analysis also employs a MELA likelihood discriminant
$\mathcal{D}_\text{bkg}^\text{kin}$ based on the \textsc{JHUGen} and \MCFM matrix element calculations
for the signal and the background, as illustrated by the inset in Fig.~\ref{fig:massfull}
and used in Ref.~\cite{Chatrchyan:2013legacy}.

\begin{figure}[hbtp]
\centering
\includegraphics[width=0.48\textwidth]{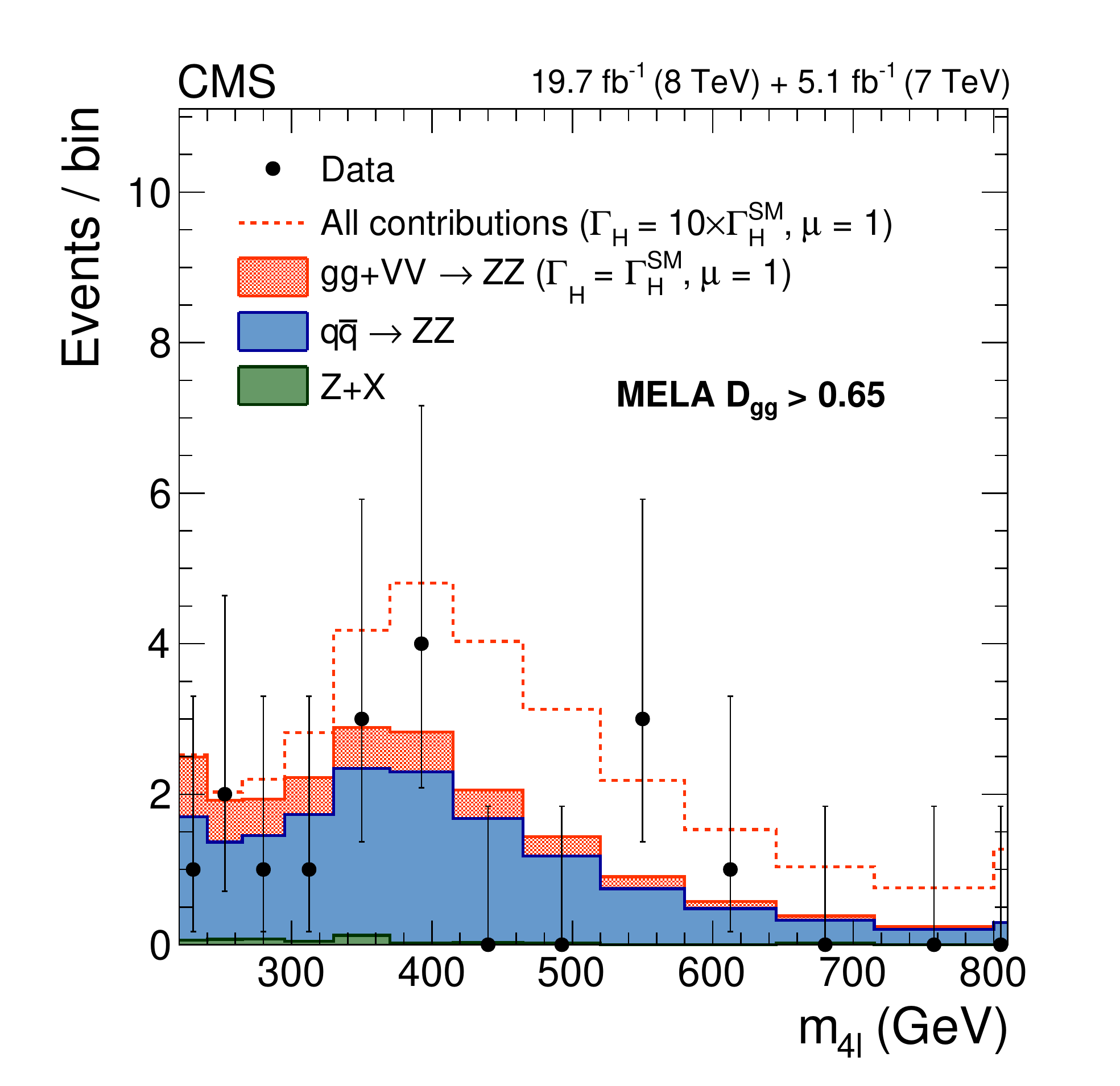}
\includegraphics[width=0.48\textwidth]{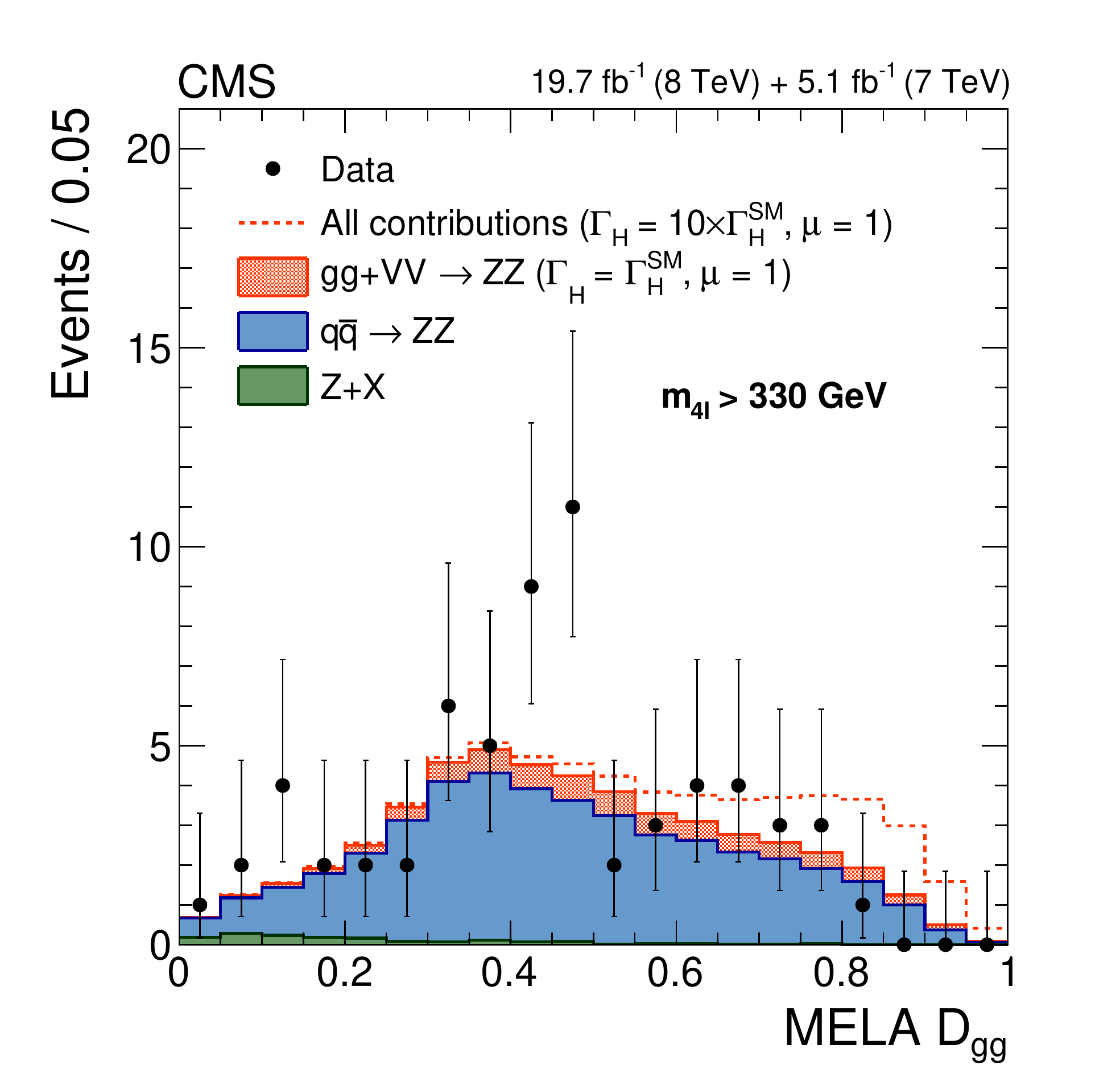}
\caption{
Distributions of (\cmsLeft) the four-lepton invariant mass after a selection requirement
on the MELA likelihood discriminant $\mathcal{D}_{\cPg\cPg} > 0.65$, and (\cmsRight) the
$\mathcal{D}_{\cPg\cPg}$ likelihood discriminant for $m_{4\ell} > 330\GeV$ in the
$4\ell$ channel. Points represent the data, filled histograms the expected
contributions from the reducible (Z+X) and $\Pq\Paq$ backgrounds, and
from the gluon fusion (gg) and vector boson fusion (VV) SM processes (including
the Higgs boson mediated contributions). The dashed line corresponds to the total
expected yield for a Higgs boson width and a squared product of the couplings scaled by a
factor 10 with respect to their SM values.
In the top plot, the bin size varies from 20 to 85\GeV and the last bin includes
all entries with masses above 800\GeV.
}
\label{fig:mss-discr}
\end{figure}

As an illustration, Fig.~\ref{fig:mss-discr}(\cmsLeft) presents the $4\ell$ invariant mass distribution
for the off-shell signal region ($m_{4\ell} > 220\GeV$) and for $\mathcal{D}_{\cPg\cPg} > 0.65$.
The expected contributions from the $\Pq \Paq \to 4\ell$ and reducible backgrounds,
as well as for the total gluon fusion (gg) and vector boson fusion (VV) contributions, including the
Higgs boson signal, are shown. The distribution of the likelihood discriminant $\mathcal{D}_{\cPg\cPg}$
for $m_{4\ell} > 330\GeV$ is shown in Fig.~\ref{fig:mss-discr}(\cmsRight), together with the expected contributions
from the SM. The expected m$_{4\ell}$ and $\mathcal{D}_{\cPg\cPg}$ distributions for the sum of all the
processes, with a Higgs boson width $\GH = 10 \times \Gamma_{\PH}^{\mathrm{SM}}$ and a relative
cross section with respect to the SM cross section equal to unity in both gluon fusion and VBF production
modes ($\mu=\mu_{\Pg\Pg\PH}=\mu_\mathrm{VBF}=1$), are also presented, showing the enhancement arising
from the scaling of the squared product of the couplings. The expected and observed event yields in
the off-shell $\cPg\cPg$-enriched region defined by $\mell \ge 330\GeV$ and $\mathcal{D}_{\cPg\cPg} >$
0.65 are reported in Table~\ref{tab:yields}.

\renewcommand{\arraystretch}{1.1}
\begin{table*}[htb]
\centering
\topcaption{Expected and observed numbers of events in the $4\ell$ and $2\ell2\nu$
channels in $\cPg\cPg$-enriched regions, defined by $\mell \ge 330\GeV$ and
$\mathcal{D}_{\cPg\cPg} >$ 0.65 ($4\ell$), and by \mt $ >350\GeV$ and $\MET > 100\GeV$ ($2\ell2\nu$).
The numbers of expected events are given separately for the gg and VBF processes,
and for a SM Higgs boson ($\GH = \Gamma_{\PH}^{\mathrm{SM}}$) and a Higgs
boson width and squared product of the couplings scaled by a
factor 10 with respect to their SM values. The unphysical
expected contributions for the signal and background components are also reported
separately, for the gg and VBF processes. For both processes, the sum of the signal
and background components differs from the total due to the negative interferences.
The quoted uncertainties include only the systematic sources.
}
\begin{tabular}{clccc}
\hline
\hline
 &                                       &  $4\ell$ & $2\ell 2\nu$ \\
\hline
(a)& Total $\cPg\cPg$ ($\GH = \Gamma_{\PH}^{\mathrm{SM}}$)  & 1.8\,$\pm 0.3$  & 9.6\,$\pm 1.5$ & \\
 & $\cPg\cPg$ Signal component ($\GH = \Gamma_{\PH}^{\mathrm{SM}}$)   & 1.3\,$\pm 0.2$  & 4.7\,$\pm 0.6$ & \\
 & $\cPg\cPg$ Background component                                          & 2.3\,$\pm 0.4$  & 10.8\,$\pm 1.7$  & \\
(b)& Total $\cPg\cPg$ ($\GH = 10 \times \Gamma_{\PH}^{\mathrm{SM}}$) & 9.9\,$\pm 1.2$  & 39.8\,$\pm5.2$ & \\
\hline
(c)& Total VBF ($\GH = \Gamma_{\PH}^{\mathrm{SM}}$)         & 0.23\,$\pm 0.01$    & 0.90\,$\pm 0.05$ & \\
 & VBF signal component ($\GH = \Gamma_{\PH}^{\mathrm{SM}}$)          & 0.11\,$\pm 0.01$    & 0.32\,$\pm 0.02$ & \\
 & VBF background component                                                  & 0.35$\pm 0.02$      & 1.22\,$\pm 0.07$ & \\
(d)& Total VBF ($\GH = 10 \times \Gamma_{\PH}^{\mathrm{SM}}$)        & 0.77\,$\pm 0.04$  & 2.40\,$\pm 0.14$ & \\
\hline
(e)& $\Pq \Paq$ background                      & 9.3\,$\pm 0.7$    & 47.6\,$\pm 4.0$ & \\
(f)& Other backgrounds                                             & 0.05\,$\pm 0.02$    & 35.1\,$\pm 4.2$ & \\
\hline
(a+c+e+f)& Total expected ($\GH = \Gamma_{\PH}^{\mathrm{SM}}$)       & 11.4\,$\pm 0.8$ & 93.2\,$\pm 6.0$ & \\
(b+d+e+f)& Total expected ($\GH = 10 \times \Gamma_{\PH}^{\mathrm{SM}}$) & 20.1\,$\pm1.4$  & 124.9\,$\pm 7.8$ & \\
\hline
& Observed                                                         & 11                & 91 & \\
\hline
\hline
\end{tabular}
\label{tab:yields}
\end{table*}

\begin{figure}[htb]
\centering
\includegraphics[width=\cmsFigWidth]{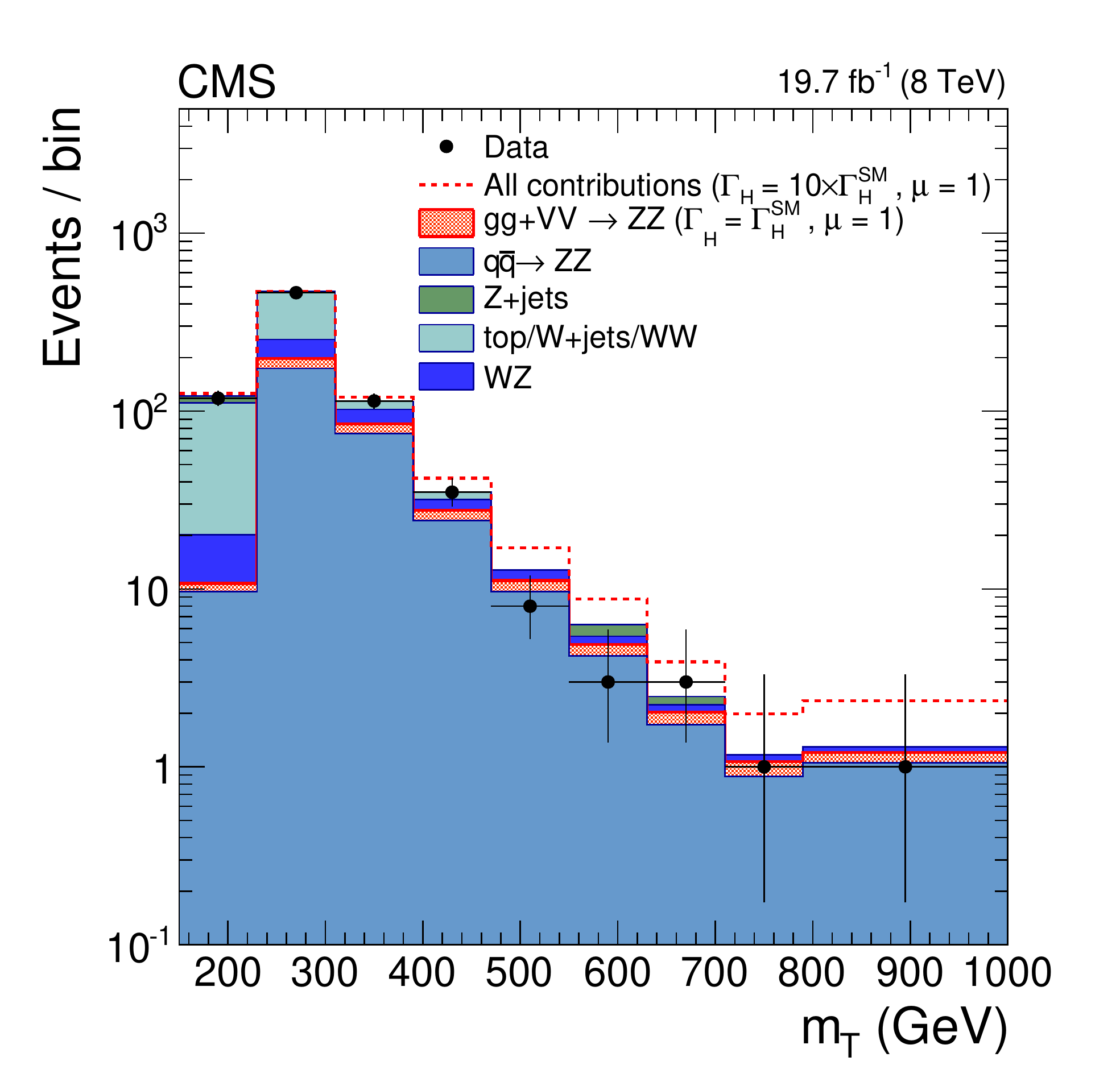}
\caption{
Distribution of the transverse mass in the $2\ell2\nu$ channel. Points represent
the data, filled histograms the expected contributions from the backgrounds, and
from the gluon fusion (gg) and vector boson fusion (VV) SM processes (including
the Higgs-mediated contributions). The dashed line corresponds to the total expected
yield for a Higgs boson width and a squared product of the couplings scaled by a
factor 10 with respect to their SM values.
The bin size varies from 80 to 210\GeV and the last bin includes all entries with
transverse masses above 1\TeV.}
\label{fig:mtfit}
\end{figure}

The $2\ell 2\nu$ analysis is performed on the 8\TeV data set only. The final state in the
$2\ell 2\nu$ channel is characterized by two oppositely-charged leptons of the same flavour
compatible with a $\cPZ$ boson, together with a large $\MET$ from the undetectable neutrinos.
We require $\MET > 80\GeV$. The event selection and background estimation is performed as
described in Ref.~\cite{Chatrchyan:2013zz}, with the exception that the jet categories defined
in Ref.~\cite{Chatrchyan:2013zz} are here grouped into a single category, i.e. the analysis
is performed in an inclusive way. The $\mt$ distribution in the off-shell signal region
($\mt > 180$\GeV) is shown in Fig.~\ref{fig:mtfit}. The expected and observed event yields in
a $\cPg\cPg$-enriched region defined by $\mt > 350\GeV$ and $\met > 100\GeV$ are reported
in Table~\ref{tab:yields}.

Systematic uncertainties comprise experimental uncertainties on the signal efficiency and
background yield evaluation, as well as uncertainties on the signal and background from
theoretical predictions. Since the measurement is performed in wide \mZZ\ regions, there
are sources of systematic uncertainties that only affect the total normalization and
others that affect both the normalization and the shape of the observables used in this
analysis. In the $4\ell$ final state, only the latter type of systematic uncertainty
affects the measurement of \GH, since normalization uncertainties change the on-shell
and off-shell yields by the same amount.

Among the signal uncertainties, experimental systematic uncertainties are evaluated
from observed events for the trigger efficiency (1.5\%), and combined object reconstruction,
identification and isolation efficiencies (3--4\% for muons, 5--11\% for electrons)~\cite{Chatrchyan:2013legacy}.
In the $2\ell 2\nu$ final state, the effects of the lepton momentum scale (1--2\%) and jet
energy scale (1\%) are taken into account and propagated to the evaluation of \MET.
The uncertainty in the b-jet veto (1--3\%) is estimated from simulation using
correction factors for the b-tagging and b-misidentification efficiencies as measured
from the dijet and \ttbar decay control samples~\cite{Chatrchyan:2012jua}.

Theoretical uncertainties from QCD scales in the $\Pq \Paq$ background contribution are
within 4--10\% depending on \mZZ~\cite{Chatrchyan:2013legacy}. An additional uncertainty of
2--6\% is included to account for missing higher order contributions with respect to a full NLO
QCD and NLO EW evaluation. The systematic uncertainty
in the normalization of the reducible backgrounds is evaluated following the methods
described in Refs.~\cite{Chatrchyan:2013legacy,Chatrchyan:2013zz}. In the $2\ell 2\nu$
channel, for which these contributions are not negligible at high mass, the estimation
from control samples for the $\cPZ$+jets and for the sum of the $\ttbar$, $\cPqt\PW$
and $\PW\PW$ contributions leads to uncertainties of 25\% and 15\% in the respective background
yields. Theoretical uncertainties in the high mass contribution from the gluon-induced
processes, which affect both the normalization and the shape, are especially important
in this analysis (in particular for the signal and interference contributions that
are scaled by large factors). However, these uncertainties partially cancel when
measuring simultaneously the yield from the same process in the on-shell signal region. The
remaining \mZZ-dependent uncertainties in the QCD renormalization and factorization
scales are derived using the K factor variations from Ref.~\cite{Passarino:1312.2397v1},
corresponding to a factor of two up or down from the nominal $m_{\cPZ\cPZ}/2$ values,
and amount to 2--4\%. For the $\cPg\cPg \to \Zo\Zo$ continuum background production, we assign
a 10\% additional uncertainty on the K factor, following Ref.~\cite{Bonvini:1304.3053} and
taking into account the different mass ranges and selections on the specific final state. This
uncertainty also affects the interference with the signal. The PDF uncertainties are estimated
following Refs.~\cite{PDF4LHC1,PDF4LHC2} by changing the NLO PDF set from MSTW2008 to CT10~\cite{Lai:2010vv}
and NNPDF2.1~\cite{nnpdf}, and the residual contribution is about 1\%. For the VBF processes,
no significant \mZZ-dependence is found regarding the QCD scales and PDF uncertainties, which
are in general much smaller than for the gluon fusion processes~\cite{LHCHiggsCrossSectionWorkingGroup:2011ti,Heinemeyer:2013tqa}.
In the $2\ell 2\nu$ final state, additional uncertainties on the yield arising from the theoretical
description of the parton shower and underlying event are taken into account (6\%).

We perform a simultaneous unbinned maximum likelihood fit of a signal-plus-background model
to the measured distributions in the $4\ell$ and $2\ell2\nu$ channels. In the $4\ell$ channel
the analysis is performed in the on-shell and off-shell signal regions defined above. In the
on-shell region, a three-dimensional distribution
$\vec{x}= (m_{4\ell}, \mathcal{D}_\text{bkg}^\text{kin}, \pt^{4\ell}\ \text{or}\ \mathcal{D}_\text{jet})$
is analyzed, following the methodology described in Ref.~\cite{Chatrchyan:2013legacy}, where the
quantity $\mathcal{D}_\text{jet}$ is a discriminant used to separate VBF from gluon fusion production.
In the off-shell region, a two-dimensional distribution $\vec{x}=(m_{4\ell}, \mathcal{D}_{\cPg\cPg})$
is analyzed. In the $2\ell2\nu$ channel, only the off-shell Higgs boson production is analyzed,
using the $\vec{x}=m_\mathrm{T}$ distribution.

The probability distribution functions are built using the full detector simulation or data control
regions, and are defined for the signal, the background, or the interference between the two contributions,
$\mathcal{P}_\text{sig}$, $\mathcal{P}_\text{bkg}$, or $\mathcal{P}_{\text{int}}$, respectively, as a function of
the observables $\vec{x}$ discussed above. Several production mechanisms are considered for the signal and
the background, such as gluon fusion (gg), VBF, and quark-antiquark annihilation ($\Pq\Paq$).
The total probability distribution function for the off-shell region includes the interference of two
contributions in each production process:
\ifthenelse{\boolean{cms@external}}{
\begin{multline}
 \mathcal{P}_\text{tot}^\text{off-shell}(\vec{x})  =
\Big[
\mu_{\Pg\Pg\PH}\times(\GH/\GZ) \times \mathcal{P}^{\Pg\Pg}_\text{sig}(\vec{x})
\\+  \sqrt{\mu_{\Pg\Pg\PH}\times(\GH/\GZ)} \times  \mathcal{P}^{\Pg\Pg}_\text{int}(\vec{x})+ \mathcal{P}^{\Pg\Pg}_\text{bkg}(\vec{x})
\Big]\\
  +
\Big[
\mu_\mathrm{VBF}\times(\GH/\GZ) \times \mathcal{P}^\mathrm{VBF}_\text{sig}(\vec{x})\\ +  \sqrt{\mu_\mathrm{VBF} \times(\GH/\GZ) } \times  \mathcal{P}^\mathrm{VBF}_\text{int}(\vec{x}) + \mathcal{P}^\mathrm{VBF}_\text{bkg}(\vec{x})
\Big]
+
\mathcal{P}^{\Pq\Paq}_\text{bkg}(\vec{x}) + \ldots
\label{eq:pdf-prob-vbf}
\end{multline}}{
\begin{equation}\begin{aligned}
 \mathcal{P}_\text{tot}^\text{off-shell}(\vec{x})  =&
\left[
\mu_{\Pg\Pg\PH}\times(\GH/\GZ) \times \mathcal{P}^{\Pg\Pg}_\text{sig}(\vec{x}) +  \sqrt{\mu_{\Pg\Pg\PH}\times(\GH/\GZ)} \times  \mathcal{P}^{\Pg\Pg}_\text{int}(\vec{x}) + \mathcal{P}^{\Pg\Pg}_\text{bkg}(\vec{x})
\right]\\
 & +
\left[
\mu_\mathrm{VBF}\times(\GH/\GZ) \times \mathcal{P}^\mathrm{VBF}_\text{sig}(\vec{x}) +  \sqrt{\mu_\mathrm{VBF} \times(\GH/\GZ) } \times  \mathcal{P}^\mathrm{VBF}_\text{int}(\vec{x}) + \mathcal{P}^\mathrm{VBF}_\text{bkg}(\vec{x})
\right] \\
& +
\mathcal{P}^{\Pq\Paq}_\text{bkg}(\vec{x}) + \ldots
\label{eq:pdf-prob-vbf}
\end{aligned}
\end{equation}
}
The list of background processes is extended beyond those quoted depending on the final state
(Z+X, top, W+jets, WW, WZ). The parameters $\mu_{\Pg\Pg\PH}$ and $\mu_\mathrm{VBF}$ are the scale factors
which modify the signal strength with respect to the reference parameterization in each production
mechanism independently. The parameter $(\GH/\GZ)$ is the scale factor which modifies the observed width
with respect to the $\GZ$ value used in the reference parameterization.

In the on-shell region, the parameterization includes the small contribution of the $\ttbar\PH$
and VH Higgs boson production mechanisms, which are related to the gluon fusion and VBF processes,
respectively, because either the quark or the vector boson coupling to the Higgs boson is in common among
those processes. Interference effects are negligible in the on-shell region. The total probability
distribution function for the on-shell region is written as
\ifthenelse{\boolean{cms@external}}{
\begin{multline}
 \mathcal{P}_\text{tot}^\text{on-shell}(\vec{x})  =
\mu_{\Pg\Pg\PH}\times \Big[ \mathcal{P}^{\Pg\Pg}_\text{sig}(\vec{x}) +  \mathcal{P}^{\ttbar\PH}_\text{sig}(\vec{x}) \Big]\\
+ \mu_\mathrm{VBF}  \times  \Big[  \mathcal{P}^\mathrm{VBF}_\text{sig}(\vec{x})  + \mathcal{P}^{\mathrm{V}\PH}_\text{sig}(\vec{x})  \Big]
 + \mathcal{P}^{\Pq\Paq}_\text{bkg}(\vec{x}) + \mathcal{P}^{\Pg\Pg}_\text{bkg}(\vec{x}) + \ldots
\label{eq:pdf-prob-onshell}
\end{multline}
}{
\begin{equation}\begin{aligned}
 \mathcal{P}_\text{tot}^\text{on-shell}(\vec{x})  =&
\mu_{\Pg\Pg\PH}\times \left[ \mathcal{P}^{\Pg\Pg}_\text{sig}(\vec{x}) +  \mathcal{P}^{\ttbar\PH}_\text{sig}(\vec{x}) \right]
+ \mu_\mathrm{VBF}  \times  \left[  \mathcal{P}^\mathrm{VBF}_\text{sig}(\vec{x})  + \mathcal{P}^{\mathrm{V}\PH}_\text{sig}(\vec{x})  \right]\\
& + \mathcal{P}^{\Pq\Paq}_\text{bkg}(\vec{x}) + \mathcal{P}^{\Pg\Pg}_\text{bkg}(\vec{x}) + \ldots
\label{eq:pdf-prob-onshell}\end{aligned}
\end{equation}
}
The above parameterizations in Eqs.~(\ref{eq:pdf-prob-vbf}, \ref{eq:pdf-prob-onshell}) are performed for the
tree-level HVV coupling of a scalar Higgs boson, consistent with our observations~\cite{Chatrchyan:2012br,Chatrchyan:2013legacy}.
We find that the presence of anomalous couplings in the HVV interaction would lead to enhanced off-shell
production and a more stringent constraint on the width. It is evident that the parameterization in
Eq.~(\ref{eq:pdf-prob-vbf}) relies on the modeling of the gluon fusion production with the dominant top-quark
loop, therefore no possible new particles are considered in the loop. Further discussion can also be found in
Refs.~\cite{Gainer:2014hha, Englert:2014aca, Passarino:1405.1925v1}.

The three parameters $\GH$, $\mu_{\Pg\Pg\PH}$, and $\mu_\mathrm{VBF}$ are left unconstrained in the fit.
The $\mu_{\Pg\Pg\PH}$ and $\mu_\mathrm{VBF}$ fitted values are found to be almost identical to
those obtained in Ref.~\cite{Chatrchyan:2013legacy}. Systematic uncertainties are included as nuisance
parameters and are treated according to the frequentist paradigm~\cite{ATLASCMS}. The shapes and
normalizations of the signal and of each background component are allowed to vary within their
uncertainties, and the correlations in the sources of systematic uncertainty are taken into account.

\begin{figure}[htb]
\centering
\includegraphics[width=\cmsFigWidth]{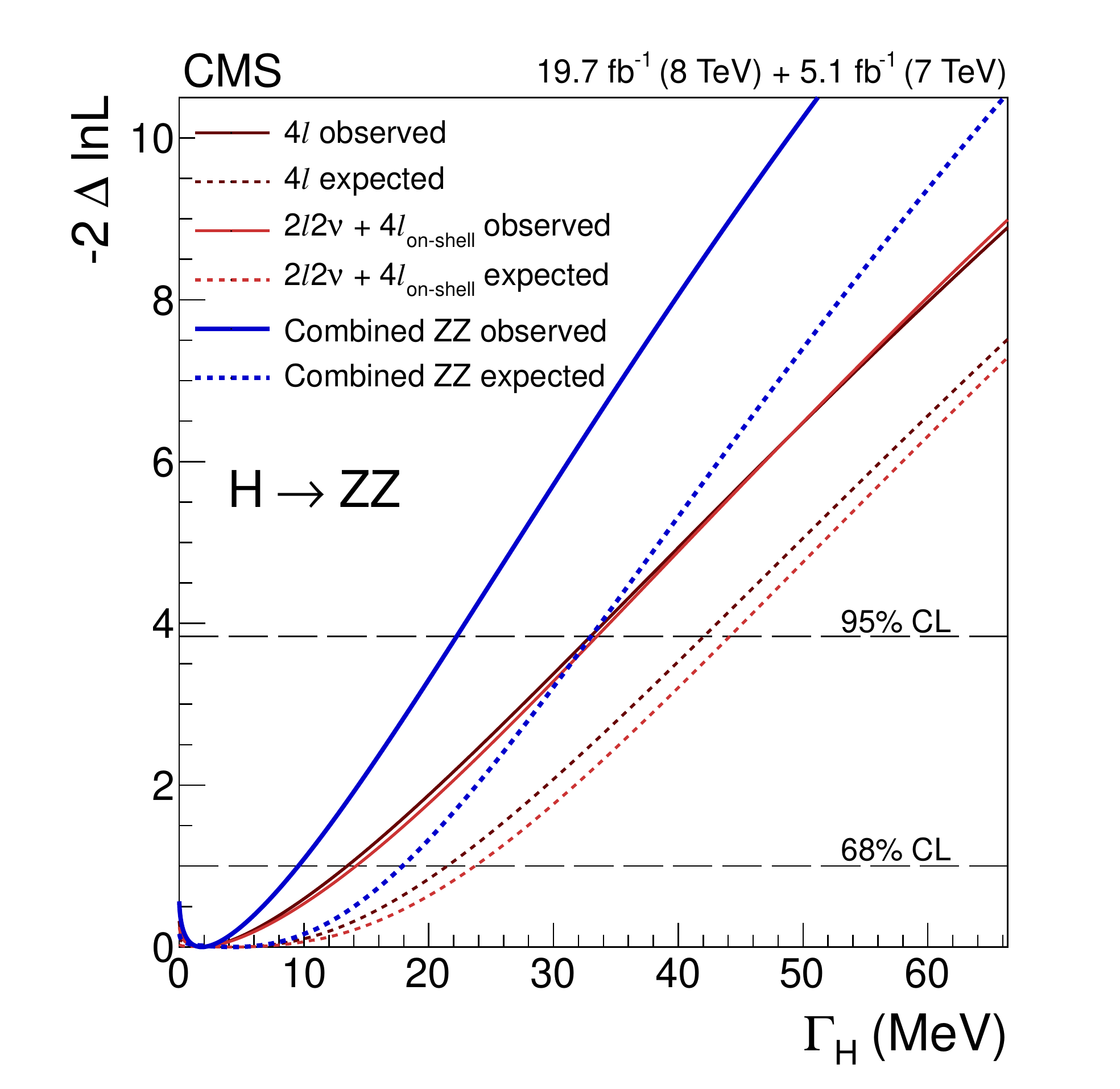}
\caption{
Scan of the negative log-likelihood, $-2 \Delta \ln\mathcal{L}$, as a
function of $\Gamma_{\PH}$ for the combined fit of the $4\ell$ and
$2\ell 2\nu$ channels (blue thick lines), for the $4\ell$ channel alone in the
off-shell and on-shell regions (dark red lines), and for the $2\ell 2\nu$
channel in the off-shell region and $4\ell$ channel in the on-shell region
(light red lines). The solid lines represent the observed values, the dotted lines
the expected values.
}
\label{fig:finalplot}
\end{figure}

The fit results are shown in Fig.~\ref{fig:finalplot} as scans of the negative
log-likelihood, $-2 \Delta \ln\mathcal{L}$, as a function of $\GH$.
Combining the two channels a limit is observed (expected) on the total width of $\Gamma_{\PH} < 22\MeV$
($33\MeV$) at a 95\% CL, which is 5.4 (8.0) times the expected value in the SM.
The best fit value and 68\% CL interval correspond to $\Gamma_{\PH} = 1.8^{+7.7}_{-1.8}\MeV$. The
result of the $4\ell$ analysis alone is an observed (expected) limit of $\Gamma_{\PH} < 33\MeV$
($42\MeV$) at a 95\% CL, which is 8.0 (10.1) times the SM value, and the result of the
analysis combining the $4\ell$ on-shell and $2\ell 2\nu$ off-shell regions is $\Gamma_{\PH} < 33\MeV$
($44\MeV$) at a 95\% CL, which is 8.1 (10.6) times the SM value.
The best fit values and 68\% CL intervals are $\Gamma_{\PH} = 1.9^{+11.7}_{-1.9}\MeV$ and
$\Gamma_{\PH} = 1.8^{+12.4}_{-1.8}\MeV$ for the $4\ell$ analysis and for the analysis combining the $4\ell$
on-shell and $2\ell 2\nu$ off-shell regions, respectively.

The expected limit for the two channels combined without including the
systematic uncertainties is $\Gamma_{\PH} < 28\MeV$
at a 95\% CL. The effect of systematic uncertainties is driven by the $2\ell 2\nu$ channel with
larger experimental uncertainties in signal efficiencies and background
estimation from control samples in data, while the result in the $4\ell$ channel is largely dominated by the statistical
uncertainty.

The statistical compatibility of the observed results with the expectation under the SM hypothesis
corresponds to a p-value of 0.24. The statistical coverage of the results obtained in the likelihood
scan has also been tested with the Feldman--Cousins approach~\cite{Feldman:1998} for the combined
analysis leading to consistent although slightly tighter constraints.
The analysis in the $4\ell$ channel has also been performed in a one-dimensional fit using either
$m_{4\ell}$ or $\mathcal{D}_{\cPg\cPg}$ and consistent results are found. The expected limit without using
the MELA likelihood discriminant $\mathcal{D}_{\cPg\cPg}$ is 40\% larger in the $4\ell$ channel.

In summary, we have presented constraints on the total Higgs boson width using its relative
on-shell and off-shell production and decay rates to four leptons or two leptons and two
neutrinos. The analysis is based on the 2011 and 2012 data sets corresponding to integrated
luminosities of \usedLumiA~at $\sqrt{s} = 7\TeV$ and  \usedLumiB at $\sqrt{s} = 8\TeV$.
The four-lepton analysis uses the measured invariant mass distribution near the peak and
above the Z-boson pair production threshold, as well as a likelihood discriminant to separate
the gluon fusion ZZ production from the $\Pq\Paq\to \cPZ\cPZ$ background,
while the two-lepton plus two-neutrino off-shell analysis relies on the transverse mass distribution.
The presented analysis determines the independent contributions of the gluon fusion and VBF production
mechanisms from the data in the on-shell region. It relies nevertheless on the knowledge of the coupling
ratios between the off-shell and on-shell production, \ie the dominance of the top quark loop in
the gluon fusion production mechanism and the absence of new particle contribution in the loop.
The presence of anomalous couplings in the HVV interaction would lead to enhanced
off-shell production and would make our constraint tighter.
The combined fit of the $4\ell$ and $2\ell 2\nu$ channels leads to an upper limit on the Higgs boson
width of $\Gamma_{\PH} < 22\MeV$ at a 95\% confidence level, which is 5.4 times the expected
width of the SM Higgs boson. This result improves by more than two orders of magnitude upon previous
experimental constraints on the new boson decay width from the direct measurement at the resonance peak.

We wish to thank our theoretician colleagues and in particular Fabrizio Caola
for providing the theoretical uncertainty in the $\cPg\cPg \to \Zo\Zo$ background
K factor, Tobias Kasprzik for providing the numerical calculations on the EW corrections
for the $\Pq \Paq \to \Zo\Zo$ background process, Giampiero Passarino
for his calculations of the $\mZZ$-dependent K factor and its variations with renormalization
and factorization scales, and Marco Zaro for checking the independence on $\mZZ$ of higher-order
corrections in VBF processes. We also gratefully acknowledge Alessandro Ballestrero, John Campbell,
Keith Ellis, Stefano Forte, Nikolas Kauer, Kirill Melnikov, and Ciaran Williams for their help in
optimizing the Monte Carlo generators for this analysis.

We congratulate our colleagues in the CERN accelerator departments for the
excellent performance of the LHC and thank the technical and administrative
staffs at CERN and at other CMS institutes for their contributions to the
success of the CMS effort. In addition, we gratefully acknowledge the computing
centres and personnel of the Worldwide LHC Computing Grid for delivering so
effectively the computing infrastructure essential to our analyses. Finally,
we acknowledge the enduring support for the construction and operation of the
LHC and the CMS detector provided by the following funding agencies: BMWF and
FWF (Austria); FNRS and FWO (Belgium); CNPq, CAPES, FAPERJ, and FAPESP (Brazil);
MES (Bulgaria); CERN; CAS, MoST, and NSFC (China); COLCIENCIAS (Colombia); MSES
and CSF (Croatia); RPF (Cyprus); MoER, SF0690030s09 and ERDF (Estonia); Academy
of Finland, MEC, and HIP (Finland); CEA and CNRS/IN2P3 (France); BMBF, DFG, and
HGF (Germany); GSRT (Greece); OTKA and NIH (Hungary); DAE and DST (India); IPM
(Iran); SFI (Ireland); INFN (Italy); NRF and WCU (Republic of Korea); LAS
(Lithuania); MOE and UM (Malaysia); CINVESTAV, CONACYT, SEP, and UASLP-FAI
(Mexico); MBIE (New Zealand); PAEC (Pakistan); MSHE and NSC (Poland); FCT
(Portugal); JINR (Dubna); MON, RosAtom, RAS and RFBR (Russia); MESTD (Serbia);
SEIDI and CPAN (Spain); Swiss Funding Agencies (Switzerland); NSC (Taipei);
ThEPCenter, IPST, STAR and NSTDA (Thailand); TUBITAK and TAEK (Turkey); NASU
and SFFR (Ukraine); STFC (United Kingdom); DOE and NSF (USA).

\bibliography{auto_generated}   

\cleardoublepage \appendix\section{The CMS Collaboration \label{app:collab}}\begin{sloppypar}\hyphenpenalty=5000\widowpenalty=500\clubpenalty=5000\textbf{Yerevan Physics Institute,  Yerevan,  Armenia}\\*[0pt]
V.~Khachatryan, A.M.~Sirunyan, A.~Tumasyan
\vskip\cmsinstskip
\textbf{Institut f\"{u}r Hochenergiephysik der OeAW,  Wien,  Austria}\\*[0pt]
W.~Adam, T.~Bergauer, M.~Dragicevic, J.~Er\"{o}, C.~Fabjan\cmsAuthorMark{1}, M.~Friedl, R.~Fr\"{u}hwirth\cmsAuthorMark{1}, V.M.~Ghete, C.~Hartl, N.~H\"{o}rmann, J.~Hrubec, M.~Jeitler\cmsAuthorMark{1}, W.~Kiesenhofer, V.~Kn\"{u}nz, M.~Krammer\cmsAuthorMark{1}, I.~Kr\"{a}tschmer, D.~Liko, I.~Mikulec, D.~Rabady\cmsAuthorMark{2}, B.~Rahbaran, H.~Rohringer, R.~Sch\"{o}fbeck, J.~Strauss, A.~Taurok, W.~Treberer-Treberspurg, W.~Waltenberger, C.-E.~Wulz\cmsAuthorMark{1}
% ...
\vskip\cmsinstskip
% ...
1:~~Also at Vienna University of Technology, Vienna, Austria\\
2:~~Also at CERN, European Organization for Nuclear Research, Geneva, Switzerland\\
% ...

\end{sloppypar}
\end{document}